\documentclass[prl,twocolumn,floatfix,showpacs ]{revtex4-1}
\UseRawInputEncoding
\usepackage{amsmath}
\usepackage{amssymb}
\usepackage{graphicx}
\usepackage{dcolumn}
\usepackage{bm}
\usepackage[colorlinks=true,linkcolor=blue,anchorcolor=blue, citecolor=cyan,urlcolor=cyan]{hyperref}
\usepackage[mathlines]{lineno}
\usepackage{ulem}
\usepackage{epstopdf}

\begin{document}
\title{Valley  polarization in 2D tetragonal altermagnetism}
\author{San-Dong Guo$^1$}
\email{sandongyuwang@163.com}
\author{Xiao-Shu Guo$^1$}
\author{Guangzhao Wang$^2$}
\affiliation{$^{1}$School of Electronic Engineering, Xi'an University of Posts and Telecommunications, Xi'an 710121, China}
\affiliation{$^{2}$Key Laboratory of Extraordinary Bond Engineering and Advanced Materials Technology of Chongqing, School of Electronic Information Engineering, Yangtze Normal University, Chongqing 408100, China}

\begin{abstract}
 The altermagnetism caused by alternating crystal environment  provides a unique opportunity for designing new type of valley polarization.
 Here, we propose a way to realize valley  polarization in two-dimensional (2D) tetragonal altermagnetism by regulating the direction of magnetization.
The valley polarization along with spin polarization will arise when the orientation of magnetization breaks the $C_{4z}$ lattice rotational symmetry, particularly in the conventional in-plane $x$ or $y$ directions. When the direction of magnetization switches between the $x$ and $y$ direction, the valley  and spin polarizations will be reversed. This is different from the widely studied valley polarization, which occurs in out-of-plane hexagonal magnetic materials  with valley physics at -K/K point. Followed by first-principles calculations, our proposal is demonstrated
in a 2D Janus tetragonal altermagnetic $\mathrm{Fe_2MoS_2Se_2}$ monolayer with good stability but very small valley splitting of 1.6 meV. To clearly see the feasibility of our proposal, an unrealistic material $\mathrm{Ru_2MoS_2Se_2}$ is used to show large valley splitting of 90 meV.
In fact,  our proposal can be readily extended to 2D tetragonal ferromagnetic (FM) materials, for example FM $\mathrm{Fe_2I_2}$ monolayer.
Our findings can enrich the valley physics, and provide new type of valley materials.

\end{abstract}

\maketitle

\section{Introduction}
Valley, characterizing the energy extrema of band, provides a new
degree of freedom to  process information and perform  logic operations  with low power consumption and high speed, and this entity is referred to as valleytronics\cite{q1,q2,q3,q4}.  To make use of valley degree, the crucial step is to break the balance of the number of carriers between the valleys, known as
valley polarization\cite{v5,v7,v9,v10,v11}.
 Two-dimensional (2D) magnetic  materials  provide the new opportunity for the long-sought spontaneous valley polarization\cite{q10}.
 In previous most works, spontaneous valley polarization is
generally considered to occur in 2D hexagonal ferromagnetic (FM) materials with out-of-plane magnetization\cite{q11,q12,q13,q13-1,q14,q14-1,q15,q16,q17,q18}, and the valley physics is concentrated at -K/K point in the  Brillouin zone (BZ). Compared with ferromagnetism, antiferromagnetic (AFM) materials possess zero magnetic moment, which are inherently robust to external magnetic perturbation, and possess ultrafast dynamics\cite{k1,k2},
thus providing  enormous potential for device applications. In fact,  the spontaneous valley polarization can also occur in hexagonal AFM monolayers\cite{gsd1,gsd2}, but  missing spin splitting  in the band
structures prevents some interesting physical phenomena.

Altermagnetism is a recently discovered third class of collinear magnetism, which shares some properties with
AFM ordering (zero magnetic moment)  and even more with FM ordering (spin splitting)\cite{k4,k5}.   The zero-magnetization magnetic
structures of altermagnetism is special-symmetry-driven, for example rotational and mirror symmetries rather than translational and  inversion symmetries. The altermagnetic spin splitting
arises in the zero spin-orbital coupling (SOC) limit.
Achieving valley  polarization  in  altermagnetism  may be significative and interesting for constructing new type of valleytronic device.
The valley can be polarized by simply breaking the corresponding crystal
symmetry with uniaxial strain in tetragonal lattice\cite{k6,k7,k7-1,k7-2,k7-3}.
Recently, twisted altermagnetism has been proposed by introducing   a key
in-plane 2-fold rotational operation, and the symmetry of spin splitting, i.e.,
$d$-wave, $g$-wave or $i$-wave, can be realized by choosing out-of-plane rotational symmetry\cite{k8}.
The twisted altermagnetism can possess valley-layer coupling, and an out-of-plane external electric field can be used to induce valley polarization\cite{k9,k10}.
\begin{figure}
  \includegraphics[width=8cm]{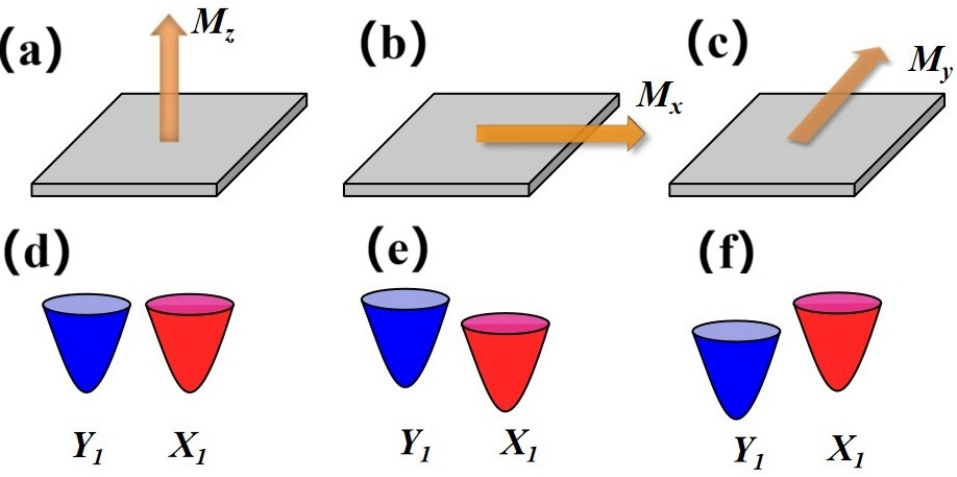}
  \caption{(Color online) (a): 2D tetragonal out-of-plane  altermagnetism possesses  equivalent valleys $Y_1$ and $X_1$ (d) with opposite spin character along $\Gamma$-Y and $\Gamma$-X lines in the BZ; (b): When the direction of magnetization is in-plane along the $x$ direction, the $Y_1$ and $X_1$ valleys become unequal (e), giving rise to valley polarization along with spin polarization; (c): When rotating the magnetization direction from $x$ to $y$, the transformation of valley and spin polarizations can be achieved (f). In (d,e,f),  the spin-up
and spin-down channels are depicted in blue and red.}\label{st}
\end{figure}

In this work, we propose  the realizability of valley  polarization in 2D tetragonal altermagnetism. The physics of the proposed way  is rooted in the combination of  SOC and in-plane magnetization. By rotating in-plane magnetization, the valley polarization can be reversed at intervals of 90$^{\circ}$.  Based on first-principles calculations, this way is validated
in a 2D Janus tetragonal altermagnetic $\mathrm{Fe_2MoS_2Se_2}$ with good stability and an unrealistic material $\mathrm{Ru_2MoS_2Se_2}$.
Therefore, the valley  design by tuning magnetization in altermagnets would
bring rich designability to valleytronics.

\begin{figure}
  \includegraphics[width=8cm]{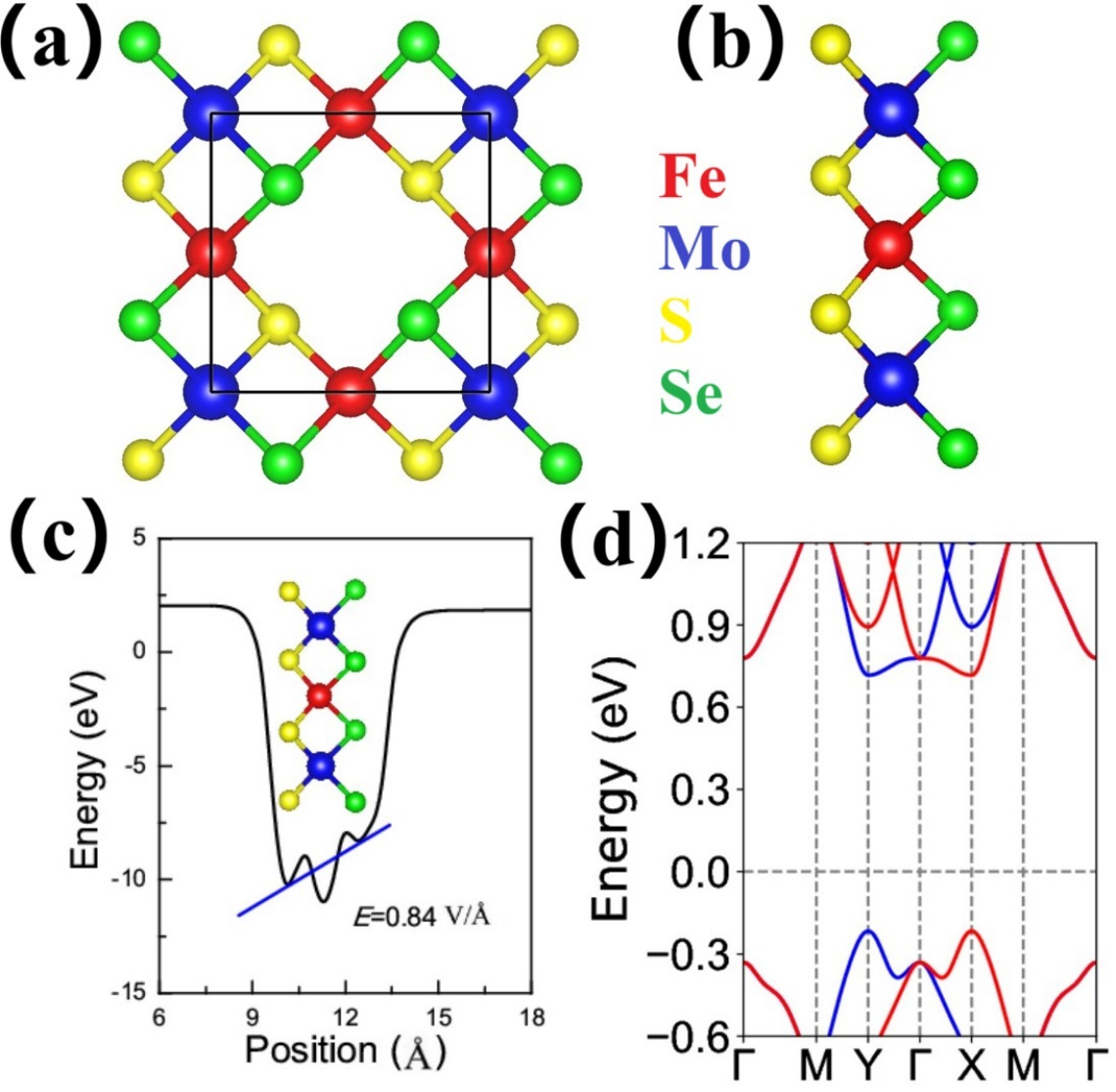}
\caption{(Color online)For $\mathrm{Fe_2MoS_2Se_2}$, (a) and (b):top and side views of the  crystal structures; (c):planar averaged electrostatic potential energy variation along $z$ direction. $E$ stands for the intrinsic polar field; (d): the energy
band structures without SOC. The spin-up
and spin-down channels are depicted in blue and red. }\label{str}
\end{figure}
\begin{figure}
  \includegraphics[width=8cm]{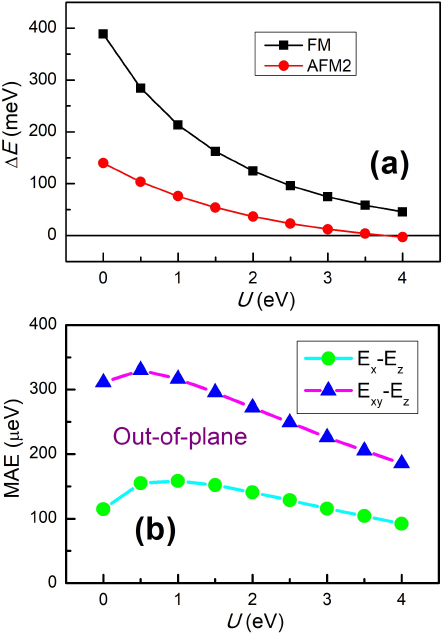}
\caption{(Color online)For $\mathrm{Fe_2MoS_2Se_2}$, the energy differences per unit cell between FM/AFM2 and AFM1 (a) and MAE (b)  as a function of $U$. }\label{emae}
\end{figure}
\begin{figure*}
  \includegraphics[width=16cm]{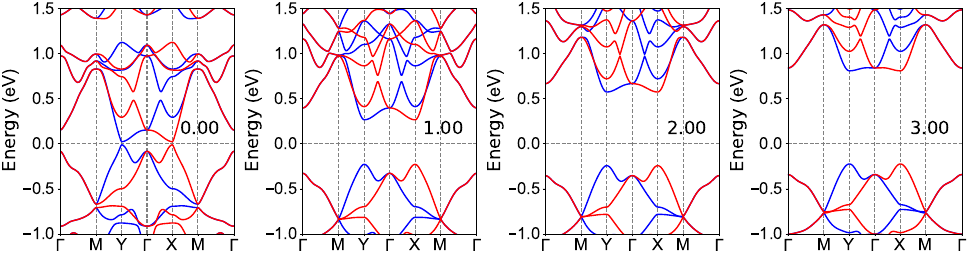}
\caption{(Color online) )For $\mathrm{Fe_2MoS_2Se_2}$, the energy band structure without SOC at representative $U$ values ($U$=0, 1, 2, 3 eV). }\label{band0}
\end{figure*}

\section{Approach and Material realization}
The spontaneous valley polarization is
generally considered to occur  in 2D hexagonal magnetic lattice (both FM and AFM systems) with  valley physics at -K/K point\cite{q10,q11,q12,q13,q13-1,q14,q14-1,q15,q16,q17,q18,gsd1,gsd2},  and the SOC and $d$ orbital characteristics play dominated role for generating the valley polarization.  For $d_{x^2-y^2}$/$d_{xy}$-dominated -K/K valley, in the general  magnetization direction $M(\theta,\phi)$ [Polar angles $\theta$ and $\phi$ define the magnetization orientation], the valley splitting $\Delta E_{VS}=4\alpha cos\theta$\cite{q18}, where the $\alpha$ is SOC-related constant, and the $\theta$=0/90$^{\circ}$ means out-of-plane/in-plane magnetization direction.
The in-plane magnetization with collinear magnetic ordering forbids the existence of valley
polarization\cite{q18}. For out-of-plane case, only $d_{x^2-y^2}$/$d_{xy}$-orbital-dominated -K and K valleys show obvious valley splitting.

Here, we assume that a 2D tetragonal altermagnetism possesses two  valleys $Y_1$ and $X_1$  along $\Gamma$-Y and $\Gamma$-X lines in BZ, which have  opposite spin character. The two valleys have no valley polarization due to $C_{4z}$ lattice symmetry in absence of SOC. While the spin splitting is present, there is an absence of spin polarization due to spin-valley locking.
With SOC, the magnetization direction has important effects on electronic structures of tetragonal altermagnetism by changing  magnetic group symmetry (the lattice symmetry combined with time reversal symmetry $T$). When the system has an out-of-plane magnetization, the magnetic point group symmetry  $C_{4z}T$  forbids the existence of valley
polarization (\autoref{st} (a) and (d)). In other words, the $x$ and $y$ axes ($\Gamma$-$Y$ and $\Gamma$-$X$  directions in BZ) are equivalent.
With in-plane magnetization along $x$ direction, the $C_{4z}T$ symmetry will be broken, producing valley and spin
polarizations (\autoref{st} (b) and (e)). When in-plane magnetization direction changes from $x$ to $y$, the valley and spin polarizations will be reversed (\autoref{st} (c) and (f)).

If the magnetization is in the general direction $M(\theta,\phi)$, no valley polarization is produced, provided that the projections $M_x (90^{\circ}, 0^{\circ})$ and $M_y (90^{\circ}, 90^{\circ})$ are equivalent.
So, when $\theta$=0$^{\circ}$, 180$^{\circ}$ or $\phi$=45$^{\circ}$, 135$^{\circ}$, 225$^{\circ}$, 315$^{\circ}$, the valley polarization will disappear.
The valley splitting  can be defined as $\Delta E_{VS}=E_{Y_1}-E_{X_1}$,  which oscillates between positive and negative values at intervals of 90 $^{\circ}$ for $\phi$ (see FIG.S1 of electronic supplementary information (ESI)).

 Monolayer $\mathrm{Cr_2O_2}$,  $\mathrm{Cr_2SO}$, $\mathrm{V_2Se_2O}$,  $\mathrm{V_2SeTeO}$ and  $\mathrm{Fe_2Se_2O}$ \cite{k6,k7,k7-1,k7-2,k7-3}  can be used to prove our proposal. These monolayers are tetragonal  altermagnetism with equivalent valleys along $\Gamma$-X and $\Gamma$-Y lines  in the BZ without SOC. Instead of available tetragonal  altermagnets, we predict a new Janus tetragonal  altermagnet $\mathrm{Fe_2MoS_2Se_2}$ with good stability.
We will use  $\mathrm{Fe_2MoS_2Se_2}$  as a protype system to illustrate the proposal of in-plane magnetization induced valley polarization in 2D tetragonal altermagnetism. The first-principles calculation method is used to prove our proposal.

\section{Computational detail}
Within density functional theory (DFT)\cite{1}, the spin-polarized  first-principles calculations are carried out  within Vienna ab
initio Simulation Package (VASP)\cite{pv1,pv2,pv3} by using the projector augmented-wave (PAW) method. The exchange-correlation potential is adopted by generalized gradient
approximation  of Perdew-Burke-Ernzerhof (PBE-GGA)\cite{pbe}.
  A kinetic energy cutoff  of 500 eV,  total energy  convergence criterion of  $10^{-8}$ eV, and  force convergence criterion of 0.0001 $\mathrm{eV.{\AA}^{-1}}$ are set to obtain reliable results.
To account for electron correlation of Fe-3$d$ orbitals, a Hubbard correction is employed within the
rotationally invariant approach proposed by Dudarev et al.
The vacuum slab of
more than 16 $\mathrm{{\AA}}$ is applied to avoid the physical interactions of periodic cells.
    A 12$\times$12$\times$1 Monkhorst-Pack k-point meshes is used to sample the BZ for calculating electronic structures.
 The phonon dispersion spectrum  is obtained by the  Phonopy code\cite{pv5}   within finite displacement method by using  4$\times$4$\times$1 supercell,.
The  ab-initio molecular dynamics (AIMD) simulations  using NVT ensemble are performed   for more than
8000 fs with a time step of 1 fs by using a 3$\times$3$\times$1 supercell. The elastic stiffness tensor  $C_{ij}$   are calculated by using strain-stress relationship, which have been renormalized by   $C^{2D}_{ij}$=$L_z$$C^{3D}_{ij}$, where the $L_z$ is  the length of unit cell along $z$ direction.

\section{crystal structure and stability}
Monolayer $\mathrm{Fe_2MoS_2Se_2}$  has a square lattice structure with the space group $Cmm2$ (No.35).  As shown in \autoref{str} (a) and (b), $\mathrm{Fe_2MoS_2Se_2}$   contains three atomic sublayers with one Mo and two  Fe atoms as  middle layer and  S/Se atoms as upper and lower layers, which  can be constructed  by  replacing one of two Se  layers with S atoms in monolayer  $\mathrm{Fe_2MoSe_4}$\cite{a1}.
To determine the magnetic ground states, the FM and two AFM configurations (AFM1 and AFM2) are constructed from a $\sqrt{2}$$\times$$\sqrt{2}$$\times$1 supercell, as shown in FIG.S2 of ESI.
The AFM1 configuration can show altermagnetism, which can be reduced into  crystal primitive cell, including  two  Fe atoms with opposite spin.
 The GGA+$U$ (The $U$=2.5 eV is adopted\cite{a2}.) results show that the
AFM1 ordering is  magnetic ground state, and its energy is 96/23 meV lower than that of FM/AFM2 case. The optimized lattice constants are $a$=$b$=5.576 $\mathrm{{\AA}}$ by GGA+$U$ for AFM1 case.
The magnetic easy-axis  can be determined by magnetic anisotropy energy (MAE), which can be obtained by calculating the energy
difference of the magnetization orientation along the (100)/(110)
and (001) cases.
Calculated results show that the MAE is 129/249 $\mathrm{\mu eV}$/Fe,  indicating that the easy-axis of $\mathrm{Fe_2MoS_2Se_2}$  is out-of-plane.
\begin{figure*}
  \includegraphics[width=16cm]{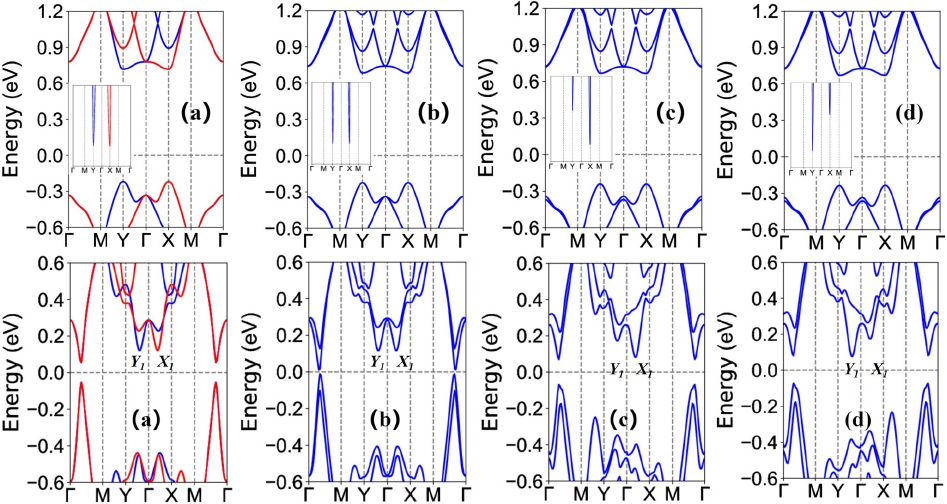}
  \caption{(Color online)For $\mathrm{Fe_2MoS_2Se_2}$ (top) and  $\mathrm{Ru_2MoS_2Se_2}$  (bottom), the energy band structures  without SOC (a), and  with SOC (b, c, d) for magnetization direction along the positive $z$, positive $x$, and positive $y$ direction, respectively.  In (a), the blue (red) lines represent the band structure in the spin-up (spin-down) direction.  For top plane, the insets show the enlarged portion of conduction bands near the Fermi energy level. }\label{band}
\end{figure*}

To explore the stability of $\mathrm{Fe_2MoS_2Se_2}$,
the phonon dispersion,  AIMD  and elastic constants are performed by using GGA+$U$ for AFM1 case.
According to FIG.S3 of ESI, the phonon
dispersions show no obvious imaginary frequencies,  indicating the dynamic stability of $\mathrm{Fe_2MoS_2Se_2}$.
To corroborate the thermal stability, the  total
energy as a function of simulation time is shown in FIG.S4 of ESI using AIMD at 300 K.
It is suggested that the thermal-induced  energy fluctuations and  changes in geometry are small, indicating that the $\mathrm{Fe_2MoS_2Se_2}$ has good thermal stability at room temperature. For $\mathrm{Fe_2MoS_2Se_2}$, the independent elastic constants are $C_{11}$=48.67  $\mathrm{Nm^{-1}}$, $C_{12}$=3.88 $\mathrm{Nm^{-1}}$ and  $C_{66}$=6.35 $\mathrm{Nm^{-1}}$,  which  satisfy the  Born  criteria of mechanical stability: $C_{11}>0$, $C_{66}>0$, $C_{11}-C_{12}>0$,  confirming  its mechanical stability.

\section{Valley polarization with in-plane magnetization}
Janus  engineering cannot destroy altermagnetism of $\mathrm{Cr_2O_2}$ and $\mathrm{V_2Se_2O}$\cite{k7,k7-2}.
Monolayer $\mathrm{Fe_2MoS_4}$ and $\mathrm{Fe_2MoSe_4}$ possess altermagnetism\cite{a1}, and Janus $\mathrm{Fe_2MoS_2Se_2}$ should also have altermagnetism. However, the Janus structure can induce an inherent electric field with the magnitude
of about  0.84 $\mathrm{V/{\AA}}$ for $\mathrm{Fe_2MoS_2Se_2}$, which can be estimated by planar average of the electrostatic potential energy along out-of-plane direction (see \autoref{str} (c)).  The built-in electric field has important effects on the electronic structures,  as we shall see in a while.
This Mo atom  leads to that two Fe atoms have the same surrounding  atomic arrangement of Mo  as a segment with different orientations, which plays a crucial role in generating the altermagnetism. If the Mo atom is removed, this altermagnetism will disappear.
In other words, the two Fe-atom sublattices  are related by $M_{xy}$ mirror or $C_{4z}$  rotational symmetry instead of any translation operation, giving rise to altermagnetism.

Without SOC, the energy band structures of $\mathrm{Fe_2MoS_2Se_2}$ is plotted in \autoref{str} (d), which shows two valleys at $X$ and $Y$ high-symmetry points for both conduction and valence bands. Due to [$C_2$$\parallel$$C_{4z}$] (The $C_2$ is the two-fold rotation perpendicular to
the spin axis.), no valley and spin polarizations can be observed.
 States around $X$ and $Y$ points are dominated by two different Fe atoms with opposite spin, producing altermagnetism  and  spin-valley locking.
 Moreover,  these bands along the $\Gamma$-$X$ and $\Gamma$-$Y$ show opposite spin character, implying $d$-wave altermagnetism\cite{k6,k7,a1}.
 The  magnetic moments of  two Fe atoms are strictly equal in size and opposite in direction, and they are 2.94 $\mu_B$ and -2.94 $\mu_B$, respectively. The total magnetic moment of $\mathrm{Fe_2MoS_2Se_2}$  is strictly 0.00 $\mu_B$.  These are confirmed by  $M_{xy}$ mirror or $C_{4z}$  rotational symmetry.

Here, we also consider the electronic correlation effects on physical properties of $\mathrm{Fe_2MoS_2Se_2}$  by changing $U$ value.
It is found that the lattice constants $a$ increases with increasing $U$ (0-4 eV), and  the magnitude of the change  is  about 0.12 $\mathrm{{\AA}}$ from 0 eV to 4 eV.  According to \autoref{emae} (a),
when $U$$<$3.7 eV,  $\mathrm{Fe_2MoS_2Se_2}$ always possesses AFM1 ground state.
The AFM2 ordering becomes ground state with $U$ being larger than 3.7 eV, and the $\mathrm{Fe_2MoS_2Se_2}$  will lose altermagnetism.
According to \autoref{emae} (b), the easy-axis of $\mathrm{Fe_2MoS_2Se_2}$  is always out-of-plane within considered $U$ range.
The evolutions of energy  band
structures and  energy bandgap as a function of $U$ are  are plotted in \autoref{band0} and FIG.S5 of ESI, respectively.
With increasing $U$, the main characteristics of energy band structures remain unchanged, and only the band gap increases.
 Compared with spin-gapless $\mathrm{Fe_2MoS_4}$ and $\mathrm{Fe_2MoSe_4}$ (Weyl semimetal) with $U$=0 eV\cite{a1},   $\mathrm{Fe_2MoS_2Se_2}$ is  a direct band gap semiconductor with gap value of 34 meV, which may be due to  inherent electric field, not the effects of lattice constants (The $a$ of $\mathrm{Fe_2MoS_2Se_2}$ is between $a$ of $\mathrm{Fe_2MoS_4}$ and $\mathrm{Fe_2MoSe_4}$.).
\begin{figure}
  \includegraphics[width=8cm]{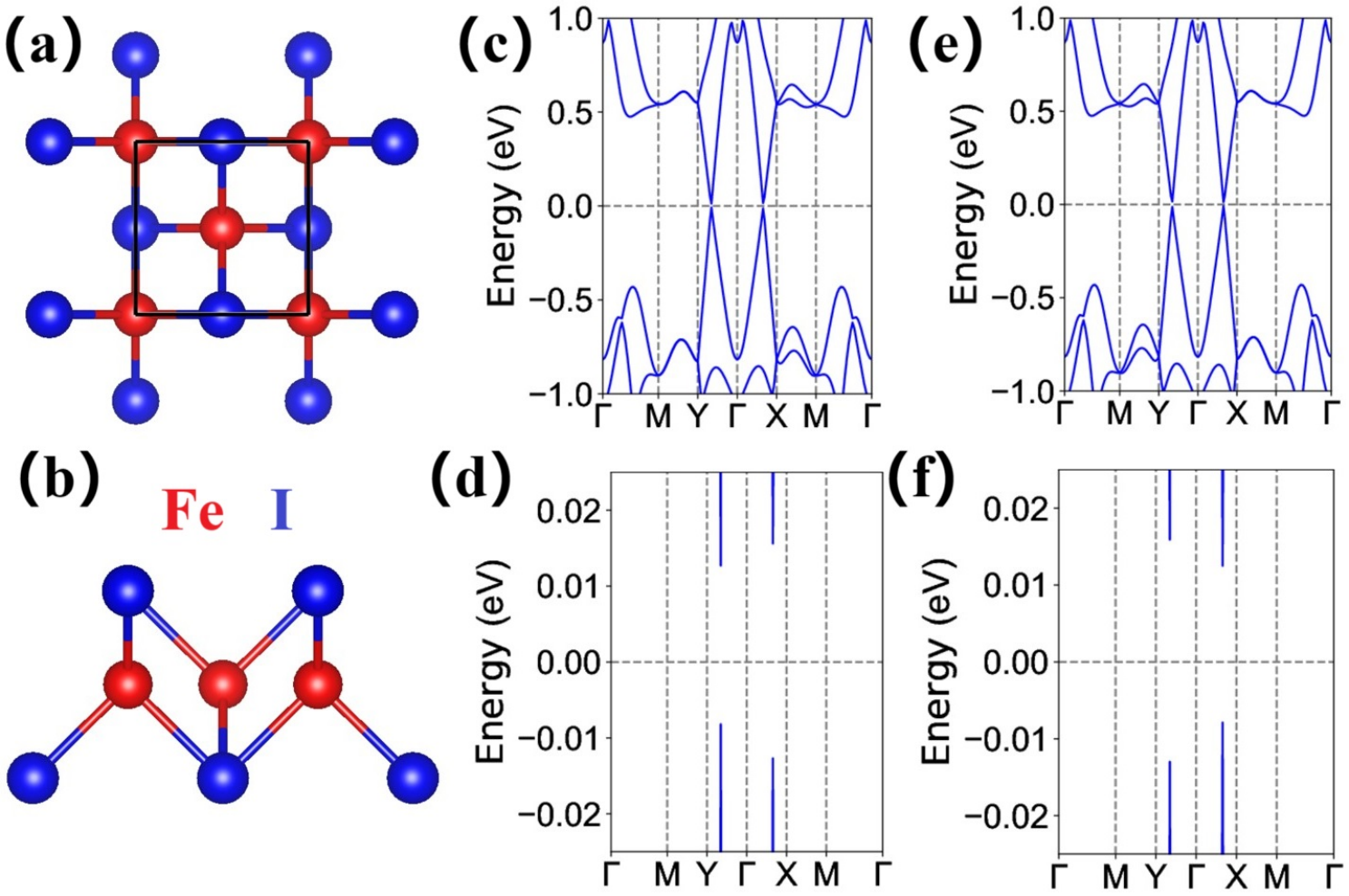}
  \caption{(Color online)For $\mathrm{Fe_2I_2}$, (a) and (b): top and side views of the  crystal structures; the energy band structures   with SOC for magnetization direction along the positive $x$ (c,d) and positive $y$ (e,f) directions, respectively.  The (d) and (f) show the enlarged portion of energy  bands near the Fermi energy level for (c) and (e). }\label{band-1}
\end{figure}

With SOC, the energy band structures and topological properties of 2D systems depend on the magnetization direction, which can affect the symmetry of a system. For 2D hexagonal magnetic lattice, only out-of-plane magnetization allows valley polarization between -K and K valleys\cite{q18}.
For $\mathrm{Fe_2I_2}$ monolayer, quantum anomalous Hall effect can only be manifested in a magnetization direction that is perpendicular to the plane\cite{a2}.
Strictly speaking, the projection of the magnetization direction possesses an out-of-plane component that facilitates the valley polarization and quantum anomalous Hall effect previously discussed.
 The energy band structures of $\mathrm{Fe_2MoS_2Se_2}$ are plotted in \autoref{band}  without SOC and  with SOC for magnetization direction  along the positive $z$, positive $x$, and positive $y$ directions.
For out-of-plane case, the  $Y$ and $X$ valleys have the same  energy, giving rise to no valley polarization due to $C_{4z}T$ symmetry.
For in-plane $x$ direction, the valley polarization can be observed due to broken $C_{4z}T$ symmetry, and the spin polarization can also occur due to spin-valley locking.
The valley splitting between $Y$ and $X$ valleys  in the conduction bands is about 1.6 meV. When the magnetization direction changes from $x$ to $y$, the valley polarization will be reversed. The energy band structures with in-plane $xy$ direction is also calculated, which shows no valley polarization (see FIG.S6), being consistent with previous discussion.

The $Y$ and $X$ valleys of $\mathrm{Fe_2MoS_2Se_2}$  are mainly from Fe character, which is a 3$d$ element with a weak SOC. To enhance valley splitting, the  mass of this magnetic atom should be increased to strengthen the SOC, for example by using 4$d$ or 5$d$ elements.
Similar strategy can also be found in hexagonal  valley materials. From  $\mathrm{FeBr_2}$ to  $\mathrm{RuBr_2}$, the valley splitting  between -K and K valleys can be significantly enhanced\cite{a3,a4}. Here, an unrealistic material $\mathrm{Ru_2MoS_2Se_2}$ is used to confirm our proposal (When $U$ is less than 3.5 eV, this spin polarization calculation converges to a non-magnetic solution. Here, we use $U$=4 eV to obtain magnetic solution, and then calculate the related energy band structures).
The energy band structures of  $\mathrm{Ru_2MoS_2Se_2}$  are shown in \autoref{band}  without SOC and  with SOC for magnetization direction  along the positive $z$, positive $x$, and positive $y$ directions. It is clearly seen that  $\mathrm{Ru_2MoS_2Se_2}$ possesses an obvious valley splitting of 90 meV between $Y_1$ and $X_1$ valleys  in the conduction bands for in-plane $x$ magnetization.  With the magnetization direction changing from $x$ to $y$, the  reversed valley polarization can be observed. Thus, the mass of the magnetic atoms determines the size of the valley splitting.

\section{Discussion and Conclusion}
 In fact, our proposed  method, analysis and results  can be readily extended to 2D tetragonal FM materials, which can possess  valley polarization with in-plane $x$ or $y$ magnetization.
 The $\mathrm{Fe_2I_2}$ monolayer has been predicted to be a stable  quantum anomalous Hall insulator with out-of-plane FM ordering\cite{a2}, and it crystallizes in the  $P4/nmm$ space group (No.129), which can be used to confirm our proposed way of realizing valley polarization.
 The crystal structures of $\mathrm{Fe_2I_2}$ are shown in \autoref{band-1} (a) and (b), which shows $C_{4z}$ symmetry.
  The optimized  equilibrium lattice constants are $a$=$b$=3.863 $\mathrm{{\AA}}$ with $U$=2.5 eV.
 The energy band structures of $\mathrm{Fe_2I_2}$ are plotted in \autoref{band-1} (c,d,e,f)   with SOC for magnetization direction  along the positive $x$ and positive $y$ directions. A valley splitting can be observed for in-plane $x$ or $y$ direction, which can be reversed by switching the direction of magnetization between $x$ and $y$. Therefore, our proposal can be extended to tetragonal magnetic systems.

In summary,   a way of realizing valley polarization in 2D materials with in-plane
magnetization is proposed in tetragonal lattice.  By rotating in-plane magnetization, the valley polarization can be reversed.
For  2D hexagonal magnetic lattice (both FM and AFM system) with  valley physics at -K/K point, the SOC plus out-of-plane magnetization is required to produce valley polarization. For our proposed 2D tetragonal magnetic lattice, the combination of SOC and in-plane magnetization is essential for generating valley polarization. The valley splitting of tetragonal  system  with 3$d$ magnetic atom is weak, and 4$d$ or 5$d$ tetragonal  system should be further searched to achieve large valley polarization.
 These explored phenomena  not only enrich the valley physics but also expand valleytronic materials.

\begin{acknowledgments}
This work is supported by Natural Science Basis Research Plan in Shaanxi Province of China  (2021JM-456). We are grateful to Shanxi Supercomputing Center of China, and the calculations were performed on TianHe-2.
\end{acknowledgments}

\end{document}